# Evaluation of Silicon-Based Photon-Counting CT for Coronary Stenosis Quantification with Realistic Coronary Artery Phantoms

Maria Jose Medrano[1]; Jed Pack[2]; Stephen Araujo[2]; Grant M. Stevens[3]; Koen Nieman[1,4]; Ge Wang[5]; Bruno De Man[2]; Adam S. Wang[1]

[1]Department of Radiology, Stanford University School of Medicine, Stanford University, Stanford, California, USA

[2]GE HealthCare Technology & Innovation Center, Niskayuna, New York, USA

[3]GE HealthCare, Waukesha, Wisconsin, USA

[4]Department of Medicine (Cardiovascular Medicine), Stanford University School of Medicine, Stanford University, Stanford, California, USA

[5]Biomedical Imaging Center, Center for Biotechnology and Interdisciplinary Research, Department of Biomedical Engineering, Rensselaer Polytechnic Institute, Troy, New York, USA

---

**Objective:** To quantify the impact of high-resolution deep silicon photon-counting CT (dSi-PCCT) on task-based coronary stenosis quantification in anatomically realistic calcified coronary artery phantoms using Micro-CT as ground truth.

**Approach:** Twelve vessel sections representing four calcification geometries (Type I–IV) and three luminal iodine concentrations (10, 15, and 20 mg/mL) were scanned under static conditions using conventional energy-integrating detector CT (EID-CT), dSi-PCCT, and Micro-CT. Images were rigidly registered to Micro-CT and segmented using an automated threshold-based pipeline. The primary analysis compared longitudinal profiles of Micro-CT-referenced percent area stenosis and segmented vessel area. A secondary analysis evaluated ellipse-derived percent area stenosis and percent vessel-area deviation at the maximum-calcification cross-section. Differences between modalities were assessed using paired block permutation and Wilcoxon signed-rank tests.

**Main results:** In the primary analysis, dSi-PCCT reduced whole-profile mean absolute error (MAE) in Micro-CT-referenced percent area stenosis from 3.10% with EID-CT to 1.62% ($p = 0.027$) and reduced segmented vessel-area MAE from 0.55 to 0.31 mm² ($p = 0.001$). In the secondary analysis, absolute deviations in ellipse-derived percent area stenosis ranged from 0.1% to 6.5% for dSi-PCCT and from 0.8% to 24.2% for EID-CT ($p < 0.001$). The greatest differences were observed at the lowest iodine concentration and for Type IV calcifications. Percent vessel-area deviation was also significantly lower with dSi-PCCT than with EID-CT ($p < 0.001$).

**Significance:** Under static, resolution-optimized imaging conditions, dSi-PCCT reduced errors in coronary vessel delineation and task-based percent area stenosis quantification relative to EID-CT. These findings establish a quantitative baseline for further evaluation in dynamic phantom and clinical CCTA studies.

## 1. INTRODUCTION

Coronary computed tomography angiography (CCTA) is recommended as a first-line imaging test for evaluating coronary artery disease (CAD) in low-to-intermediate risk patients (Gulati et al., 2021). However, CAD detection in CCTA exams remains limited for high-risk patients with dense calcifications and stents, where it is not routinely recommended (Gulati *et al.*, 2021; Narula *et al.*, 2021). This limitation is largely driven by calcium blooming artifacts and blurring-induced contrast reduction in narrow coronary segments, which can obscure lumen boundaries and lead to overestimation of vascular stenosis (Kalisz *et al.*, 2016; Song *et al.*, 2019; Mergen *et al.*, 2022). In most clinically available scanners, these partial volume artifacts are strongly influenced by the limited spatial resolution of energy-integrating detectors (Latina *et al.*, 2021; Hagar *et al.*, 2023; Koons *et al.*, 2024). Recently, photon-counting CT (PCCT) has substantially expanded the achievable spatial resolution of clinical CT imaging while also improving x-ray dose efficiency. Prior studies comparing PCCT with EID-CT have demonstrated reduced blooming artifacts from heavy calcifications and coronary stents (Koons *et al.*, 2024). Additional studies have also reported improvements in coronary artery stenosis detection, quantification, and classification relative to EID-CT (Boccalini *et al.*, 2022; Verelst *et al.*, 2023, 2025). However, establishing a reliable reference standard for stenosis quantification in these comparative studies remains challenging.

In some studies, only direct comparisons between EID-CT and PCCT were available without an independent ground truth reference (Koons *et al.*, 2024). In others, stenosis measurements were compared against invasive coronary angiography (ICA), the clinical reference standard for luminal stenosis (Boussoussou *et al.*, 2025; Fahrni *et al.*, 2025; Kotronias *et al.*, 2025). Because ICA is based on two-dimensional projection images, whereas CT provides three-dimensional volumetric data, the two modalities may derive reference lumen dimensions differently; therefore, close quantitative agreement in stenosis measurements may not be expected (Fahrni *et al.*, 2025). Coronary artery phantoms with known ground truth measurements offer a practical alternative for quantitative performance evaluation. However, prior phantom studies lacked the anatomical realism and plaque diversity of true coronary disease (Verelst *et al.*, 2023; Holmes *et al.*, 2024), whereas prior ex vivo studies in cadaveric coronary specimens, lacking an opacified lumen, quantified only calcification volume rather than luminal stenosis (Sandstedt *et al.*, 2021; Marsh *et al.*, 2023). Furthermore, coronary stenosis quantification accuracy remains underexplored for other PCCT systems that do not use cadmium telluride (CdTe) detectors.

Recently, a novel photon-counting CT system based on deep silicon photon-counting detectors (dSi-PCCT) received FDA 510(k) clearance. The goal of this study was to evaluate the impact of high-resolution imaging on coronary stenosis quantification using a previously developed set of anatomically realistic coronary artery phantoms representing multiple calcification geometries and iodine concentrations (Pack *et al.*, 2024). To enable direct quantitative evaluation, the phantoms were scanned using conventional EID-CT, deep silicon PCCT, and a Micro-CT system that provided ground-truth vessel and lumen measurements. This experimental framework enabled a controlled, ground-truth-referenced comparison of stenosis quantification using the evaluated system-specific high-resolution protocols. Through this study, we aim to establish a baseline for the stenosis quantification improvements achievable through the inherent high-resolution imaging capabilities of deep silicon PCCT. Such a baseline would help guide the interpretation of future clinical studies and support the development of complementary imaging approaches, including motion correction and advanced reconstruction techniques.

## 2. METHODS AND MATERIALS

A set of realistic coronary artery phantoms derived from clinical coronary CT angiography (CCTA) examinations was scanned using three CT systems: a conventional energy-integrating detector CT (EID-CT), which served as the comparative baseline; a deep silicon-based photon-counting CT (dSi-PCCT) evaluated in this study; and a Micro-CT, which provided high-resolution reference images for stenosis assessment in the coronary artery phantoms. To isolate the effects of intrinsic spatial resolution from motion-related artifacts, all scans were performed under static conditions. Lumen and calcification measurements for the coronary artery phantoms were obtained using an image registration and segmentation pipeline. Two quantitative metrics of stenosis grading performance were subsequently evaluated across four representative calcification geometries and three iodine concentrations. The following sections describe the anatomically realistic coronary artery phantoms used in this study (Sect. 2.1), the acquisition protocols for the EID-CT, dSi-PCCT, and Micro-CT systems (Sect. 2.2), and the image processing pipeline and quantitative metrics used to assess task-based performance (Sect. 2.3).

### *2.1. Design and Fabrication of Coronary Artery Phantom*

The anatomically realistic coronary artery phantoms characterized in this study were originally introduced by Pack et al. (Pack *et al.*, 2024). Phantom geometries were generated by extracting the longest continuous coronary artery segments from a set of clinical CCTA examinations. Automated centerline extraction and lumen segmentation were subsequently used to generate three-dimensional vessel meshes for each coronary artery. Three representative vessel sections spanning a range of luminal diameters (2–5 mm) and vessel curvatures were manually selected from each mesh. To satisfy manufacturing constraints, each selected section was designed to fit within a cylindrical volume measuring 7.5 cm in length and 2.5 cm in diameter. Mathematical interpolation between adjacent centerline points was then performed to constrain the vessel geometry to a single plane while preserving the overall vessel curvature and morphology. Calcification pits were subsequently incorporated into the vessel mesh to enable deposition of calcified plaque during the manufacturing process (Figure 1a). Following the calcification grading system first proposed by Qi et al. (Qi *et al.*, 2016) (Figure 1b), calcified plaque geometries were categorized into four calcification types (I–IV) according to the number of vessel-wall quadrants occupied by plaque in cross-sectional view. Calcification pits were incorporated into seven to eight regions along each vessel segment and were designed to span one, two, three, or four quadrants of the lumen circumference, with Type IV corresponding to plaque surrounding the vessel lumen. Two additional Type II subtypes were included: Type IIa, consisting of plaque occupying two consecutive quadrants, and Type IIb, consisting of two smaller plaques positioned in opposing quadrants.

Physical vessel phantoms containing calcification pits were fabricated using high-resolution 3D printing (Phantom Laboratories, Greenwich, NY, USA). A two-part rubber mold was subsequently cast around each printed vessel geometry, with the mold separation plane aligned with the final vessel centerline. After curing, the original 3D-printed vessel was removed, leaving a negative cavity corresponding to the vessel lumen geometry between the two mold halves. Following mold reassembly, a casting material composed of iodinated contrast agent mixed with liquefied plastic was injected into the mold cavity to reproduce the x-ray attenuation properties of iodinated blood. This casting procedure was repeated for different iodine concentrations of 10, 15, and 20 mg/ml iodine.

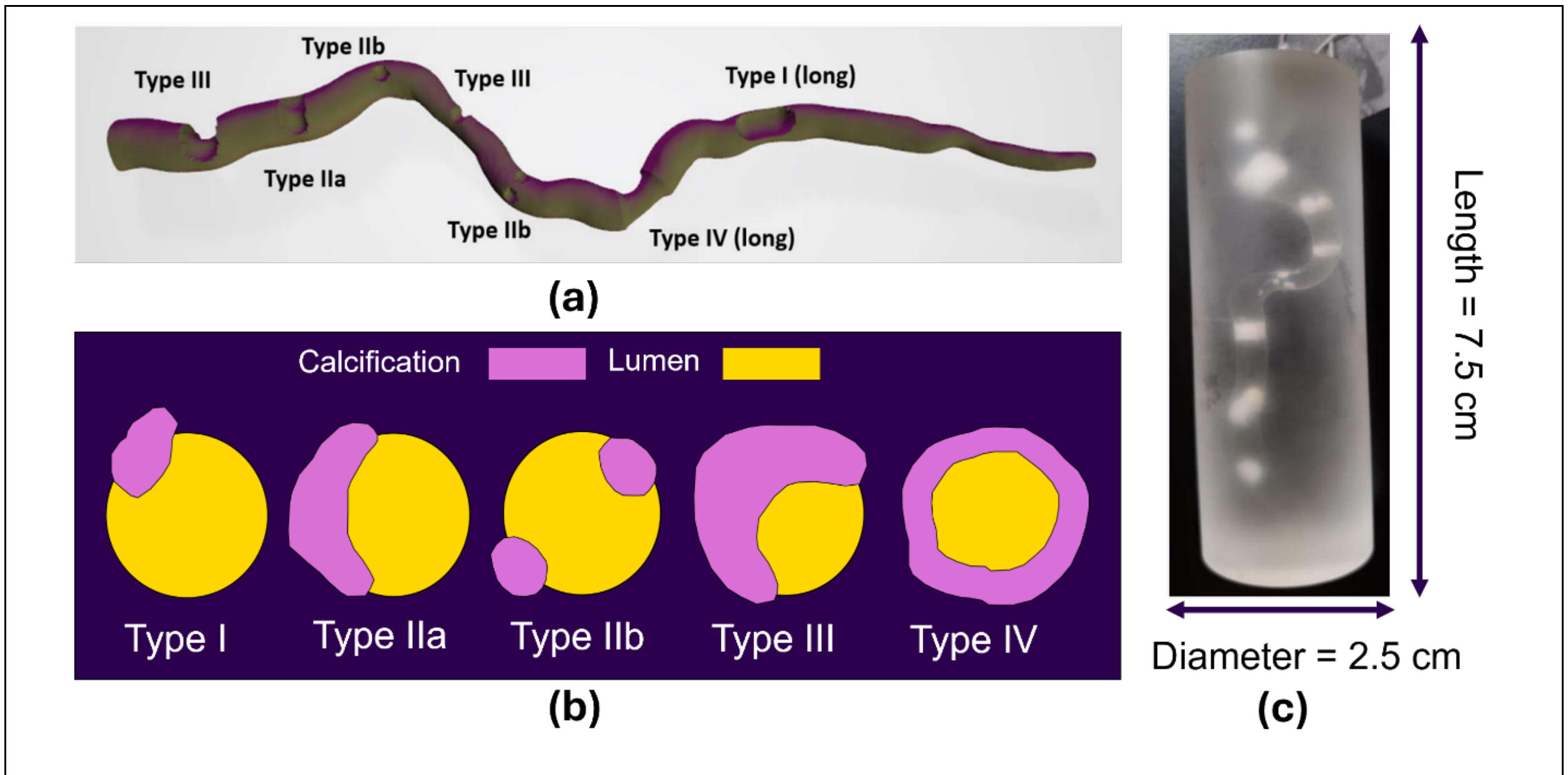

**Figure 1:** (a) Sample 3D-printed coronary artery phantom model. Pits along the vessel wall were designed to accommodate calcium-epoxy paste. (b) Representative calcified plaque geometries as defined by (Qi *et al.*, 2016), classified by the number of vessel-wall quadrants occupied by the plaque in cross-sectional view. (c) Coronary vessel phantom positioned within a cylindrical enclosure.

Two vessel replicas, one with low (10 or 15 mg/ml) and one with high (15 mg/ml or 20 mg/ml) iodine concentration, were fabricated, resulting in a total of six vessel phantoms. A calcium-epoxy paste was subsequently applied to the calcification pits, with calcium salt concentrations selected to reproduce the attenuation of clinical calcified plaque (~1100 HU). Each vessel phantom was then positioned inside a cylindrical enclosure measuring 2.5 cm in diameter and 7.5 cm in length. Finally, liquefied plastic was cast around the vessel structures to produce solid cylindrical coronary artery phantoms (Figure 1c). Further details on the design and fabrication of the coronary artery phantoms can be found in Pack et al. (Pack *et al.*, 2024).

*2.2. Data Acquisition and Image Reconstruction*

The six fabricated phantoms were scanned two at a time: the high- and low-iodine concentration versions of the three phantom geometries were secured within a 10 cm diameter water cylinder using a custom plastic support frame. The water cylinder was subsequently positioned inside a 20 cm × 30 cm cardiothoracic phantom (QRM, Moehrendorf, Germany) (Figure 2). The complete phantom assembly was scanned first on a conventional energy-integrating detector CT scanner (EID-CT; Revolution Apex, GE HealthCare, Waukesha, WI) using an axial acquisition protocol with 16 cm z-coverage (256 × 0.625 mm), 120 kVp tube potential, 420 mA tube current, and a gantry rotation time of 1.0 s. Reconstructions were performed with ASiR-V 50% and Bone kernel. The same experimental setup was subsequently scanned using a deep silicon photon-counting CT system (dSi-PCCT; Photonova Spectra, GE HealthCare, Waukesha, WI). To ensure equivalent radiation dose between systems, dSi-PCCT acquisitions were performed in axial mode using matched acquisition parameters, including 120 kVp tube potential, 420 mA tube current, 1.0 s gantry rotation time, small focal spot, and 80 mm z-collimation. Images were reconstructed using the vendor's deep-learning-based reconstruction algorithm (DL-High) in ultra-high-definition mode (UHD Ultra) with a 1024 × 1024 matrix and a field of view of 12 cm. For EID-CT and dSi-PCCT, system-specific reconstruction settings were selected to maximize the spatial-resolution performance available on each scanner for the calcification task.

High-resolution images of lumen and calcifications were acquired by scanning each of the six phantom cylinders individually using a Micro-CT system (Phoenix V|tome|x M, Baker Hughes, Hurth, Germany). These images served as ground truth reference throughout the analysis. Scans were performed in axial mode using a nano-focus x-ray tube, with two acquisitions required to cover the full 7.5 cm longitudinal extent of each phantom. Acquisition parameters included 120 kVp tube potential, 160 µA tube current, 333 ms dwell time (averaged over seven repeated acquisitions), and 1000 projection views, resulting in an approximate scan time of 45 minutes for each axial acquisition and 0.02 mm isotropic resolution. Additional scan and reconstruction parameters are summarized in Table I.

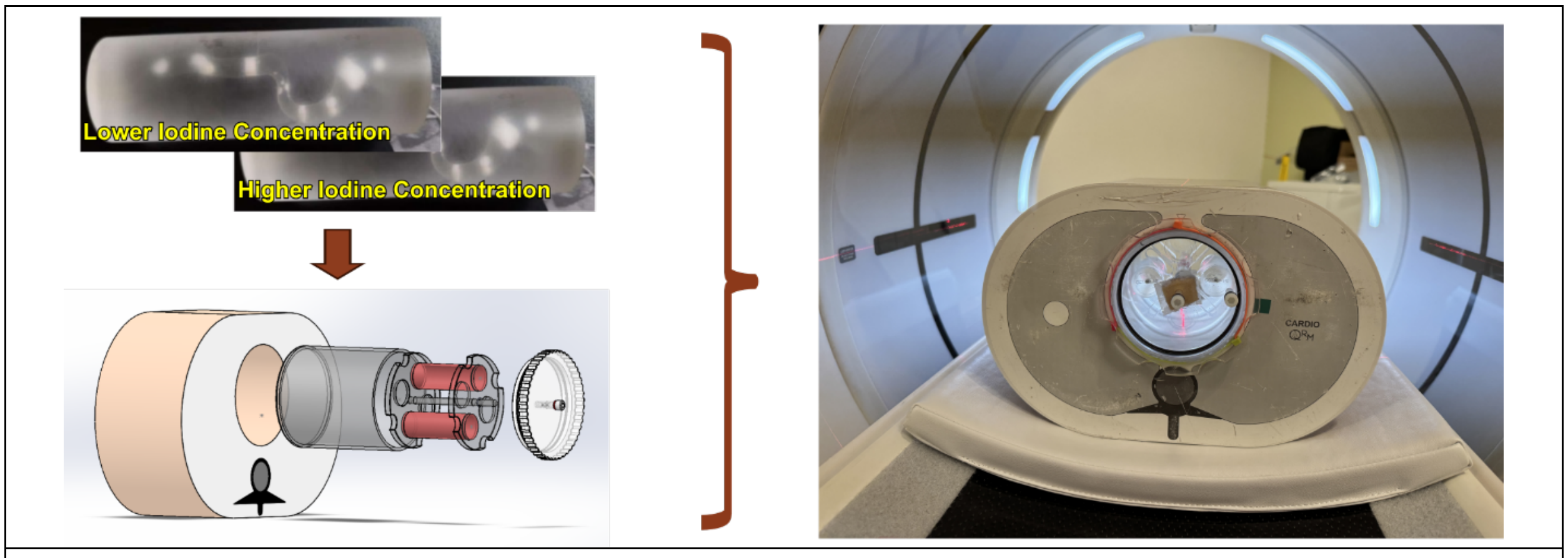

**Figure 2:** Experimental setup for scanning coronary artery phantoms using dSi-PCCT and EID-CT systems. Coronary artery phantoms were inserted into a water-filled acrylic cylinder positioned inside a QRM Thoracic Lung Phantom.

**Table I:** Acquisition and reconstruction parameters for the three CT systems in this study

| | kVp | mA | Scan Time (sec) | Focal Spot | Axial Coverage (mm) | Voxel Size (mm$^3$) | Reconstruction | Kernel |
|---|---|---|---|---|---|---|---|---|
| EID-CT | | 420 | 1.0 | S | 160 | 0.39x0.39x0.63 | ASiR-V 50% | Bone |
| dSi-PCCT | 120 | 420 | 1.0 | S | 80 | 0.12x0.12x0.41 | DL-High | UHD Ultra |
| Micro-CT | | 0.16 | 2700 (x2) | Nano | 37 (x2) | 0.02x0.02x0.02 | FBP | - |

### 2.3. Image Processing and Performance Evaluation

#### 2.3.1 Image Registration and Segmentation

A total of 12 vessel sections representing Type I-IV calcification geometries and iodine concentrations of 10, 15, and 20 mg/ml were selected from the high-resolution Micro-CT datasets. Corresponding vessel sections were identified in the EID-CT and dSi-PCCT datasets and extracted from the reconstructed volumes. All datasets were resampled to a common voxel grid $0.02 \times 0.02 \times 0.04$ mm³ using cubic B-spline interpolation. To compensate for small differences in phantom orientation and positioning among acquisitions, the cropped EID-CT and dSi-PCCT sections were rigidly registered to their corresponding Micro-CT reference volumes using SimpleITK (Lowekamp *et al.*, 2013). Registration was performed using a 3D Euler rigid transformation, Mattes mutual information as the similarity metric, and random voxel sampling. Optimization was performed using gradient descent with a learning rate of 0.5 over 40 iterations. The resulting rigid transformations were applied to the resampled CT datasets using linear interpolation to minimize additional smoothing.

Because lumen and calcification attenuation values were relatively homogeneous across phantoms, intensity-based three-class multi-Otsu thresholding ($K = 3$) was used to classify each voxel as background, lumen, or calcification to reduce reliance on manual segmentation. The highest-intensity class was identified as calcification, while the remaining two classes delineated the vessel lumen and background. Because each voxel received a single class label, the resulting lumen and calcification masks were mutually exclusive. Morphological post-processing was subsequently applied to remove isolated components through area-based filtering and to fill small gaps while preserving the spatial continuity of the calcified regions. For vessel sections containing the highest iodine concentration, a two-step multi-Otsu segmentation strategy was implemented to enable calcification segmentation in regions with reduced contrast-to-noise ratio.

#### 2.3.2 Assessment of Percent Area Stenosis and Vessel Area Deviation

Two complementary analyses were performed to evaluate stenosis quantification and vessel delineation. The primary analysis used Micro-CT as ground truth to quantify measurement accuracy along the full longitudinal extent of each calcified vessel section. A secondary analysis evaluated an ellipse-derived reference area at the slice of maximum calcification using an image-based method intended for settings in which Micro-CT ground truth is unavailable.

Clinical stenosis severity is conventionally assessed at the minimum lumen relative to a representative reference lumen, commonly estimated from adjacent proximal and distal normal segments. However, the patient-derived phantom vessels exhibited genuine longitudinal variation in cross-sectional area. Applying a single reference area across the calcified segment would therefore confound variation in vessel caliber with blooming-related measurement error. For this task-based phantom analysis, the primary analysis used the corresponding Micro-CT vessel area at each longitudinal position and evaluated the resulting full vessel profiles, whereas the secondary ellipse-derived analysis provided a local reference at the maximum-calcification cross-section.

For each imaging modality $m$ (Micro-CT, EID-CT, and dSi-PCCT), the segmented vessel mask at each longitudinal position $z$ was defined as the union of the mutually exclusive lumen and calcification masks. Accordingly, the segmented vessel area was calculated as

$$A_{vessel,m}(z) = A_{lumen,m}(z) + A_{calc,m}(z) \quad \text{Eq. 1}$$

where $A_{\text{lumen},m}(z)$ and $A_{\text{calc},m}(z)$ represent the lumen and calcification areas, respectively, for modality $m$ at longitudinal position $z$ (Figure 3a). Measurements across $z$ formed the longitudinal segmented vessel-area profile for each modality.

*Primary Micro-CT-Referenced Analysis*

For the primary analysis, the vessel area obtained from the registered high-resolution Micro-CT image, $A_{\text{vessel},\mu\text{CT}}(z)$, was used as the ground-truth reference denominator. Micro-CT-referenced percent area stenosis for each modality, denoted as $S_{m,\mu CT}(z)$, was calculated as

$$S_{m,\mu CT}(z) = 100 * \frac{A_{calc,m}(z)}{A_{vessel,\mu CT}(z)} \quad \text{Eq. 2}$$

Using the same Micro-CT denominator for EID-CT and dSi-PCCT enabled direct evaluation of differences in the segmented calcification area without introducing modality-dependent differences in the reference vessel boundary. The EID-CT and dSi-PCCT segmented vessel-area profiles, $A_{\text{vessel},m}(z)$, were also compared directly with the corresponding Micro-CT vessel-area profile to quantify errors in vessel delineation. Percent area stenosis and segmented vessel area were evaluated along the longitudinal direction to characterize measurement errors across the full calcified vessel segment.

*Secondary Ellipse-Derived Analysis*

A secondary analysis was performed at the slice of maximum calcification to evaluate a reference-area estimation method that could be used when Micro-CT ground truth is unavailable. For each modality, the slice containing the maximum calcification area was identified from the corresponding calcification mask. Following rigid registration, the datasets were longitudinally aligned by assigning the modality-specific maximum-calcification slices to a common reference position.

At the selected cross-section, an elliptical contour was generated using an automated constrained scaling procedure. The ellipse orientation and aspect ratio were initialized from the filled segmented vessel, scaled to fit within the patent lumen, and then uniformly expanded to encompass the calcification mask without exceeding the segmented vessel boundary (Figure 3b). The area enclosed by this contour, $\hat{A}_{\text{ref},m}$, represented the modality-specific ellipse-estimated reference vessel area.

Ellipse-derived percent area stenosis was calculated as

$$\hat{S}_{m,ellipse} = 100 * \frac{A_{calc,m}}{\hat{A}_{ref,m}} \qquad \text{Eq. 3}$$

Within the fitted contour, the complementary lumen area was implicitly defined as $\hat{A}_{\mathrm{ref},m} - A_{\mathrm{calc},m}$. Ellipse-derived measurements from EID-CT and dSi-PCCT were compared with the corresponding ellipse-derived Micro-CT measurements to evaluate the effect of imaging resolution when applying the same image-based reference estimation method.

Percent vessel-area deviation relative to the ellipse-estimated reference was quantified as

$$\Delta A_{m,ellipse} = 100 * \frac{\left|A_{vessel,m} - \hat{A}_{ref,m}\right|}{\hat{A}_{ref,m}} \qquad \text{Eq. 4}$$

Here, $\Delta A_{m,ellipse}$ represents the percent deviation of the segmented vessel area from the ellipse-estimated reference area. All quantities in Equations 3 and 4 were evaluated at the selected maximum calcification cross-section; therefore, dependence on $z$ is omitted.

The Micro-CT-referenced longitudinal analysis served as the primary assessment of measurement accuracy, whereas the ellipse-derived measurements provided a secondary evaluation of an image-based reference estimation method.

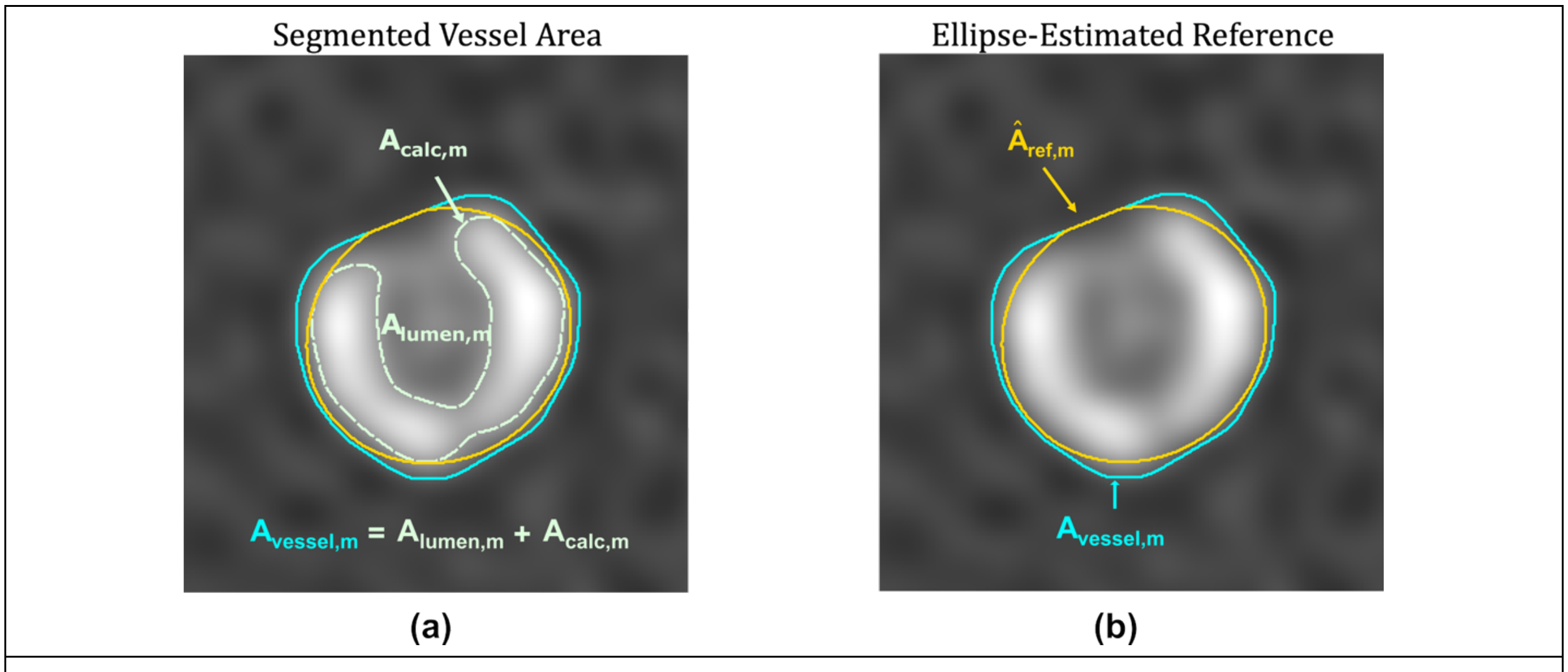

**Figure 3:** Illustration of the segmented vessel area and ellipse-derived reference measurement. (a) The segmented vessel mask (solid cyan) was defined as the union of the mutually exclusive lumen mask (not outlined) and calcification mask (dashed green), such that $A_{\mathrm{vessel},m} = A_{\mathrm{lumen},m} + A_{\mathrm{calc},m}$. The ellipse-derived reference contour is also shown in solid yellow for spatial comparison. (b) The ellipse-estimated reference area, $\hat{A}_{\mathrm{ref},m}$, was fitted to circumscribe the calcification mask while remaining within the segmented vessel boundary, $A_{\mathrm{vessel},m}$, and was used for the secondary image-based analysis.

*2.3.3 Comparative Analysis*

Statistical comparisons were performed separately for the primary Micro-CT-referenced longitudinal analysis and the secondary ellipse-derived analysis. For the primary analysis, the modality-specific profiles were longitudinally aligned at the slice of maximum calcification and linearly interpolated onto a common overlapping z-grid. For each vessel section, profile-level mean absolute error (MAE) was calculated for Micro-CT-referenced percent area stenosis and segmented vessel area. Percent area stenosis MAE was calculated as the mean absolute difference between the EID-CT or dSi-PCCT profile and the corresponding Micro-CT reference profile across all longitudinal positions. Segmented vessel-area MAE was calculated similarly by comparing each modality-specific vessel-area profile with the corresponding Micro-CT profile. Differences in profile-level MAE between dSi-PCCT and EID-CT were evaluated using a two-sided paired block permutation test, with each calcification geometry-iodine concentration combination treated as a matched block. The test statistic was based on the summed within-profile differences in MAE between EID-CT and dSi-PCCT. A null distribution was generated using 10,000 random permutations in which the modality labels were independently exchanged within each matched profile. The 12 profiles representing all calcification geometries and iodine concentrations were pooled for the overall comparison.

For the secondary analysis, absolute deviations in ellipse-derived percent area stenosis from the corresponding ellipse-derived Micro-CT measurements were calculated for EID-CT and dSi-PCCT. Differences in these deviations were evaluated using a two-sided paired Wilcoxon signed-rank test pooled across all calcification geometries and iodine concentrations (n=12 paired observations). Percent vessel-area deviation relative to the modality-specific ellipse-estimated reference was compared directly between dSi-PCCT and EID-CT using a separate two-sided paired Wilcoxon signed-rank test. A p-value less than 0.05 was considered statistically significant.

## 3. RESULTS

Figure 4 shows representative images of Micro-CT, dSi-PCCT, and EID-CT for Type I–IV calcification geometries across iodine concentrations of 10, 15, and 20 mg/ml. Images were selected at the slice of maximum calcification area. Micro-CT demonstrated the clearest delineation of calcification relative to the lumen and background across all calcification types and iodine concentrations, followed by dSi-PCCT, and then EID-CT.

### *3.1. Primary Micro-CT-Referenced Analysis*

Figure 5 shows representative Micro-CT grayscale images overlaid with segmented vessel contours and calcification masks derived from EID-CT and dSi-PCCT at the slice of maximum calcification for coronary phantoms containing 10 mg/mL iodine. Each row also presents longitudinal profiles of Micro-CT-referenced percent area stenosis and segmented vessel area for each modality. EID-CT consistently overestimated percent area stenosis and segmented vessel area relative to the Micro-CT reference compared with dSi-PCCT, with the largest overestimation observed for the Type IV calcification geometry. Profile-level MAE in Micro-CT-referenced percent area stenosis ranged from 0.56% to 1.39% for dSi-PCCT and from 1.09% to 8.89% for EID-CT across Type I–IV geometries. Similarly, segmented vessel-area MAE ranged from 0.20 to 0.54 $mm^2$ for dSi-PCCT and from 0.42 to 1.13 $mm^2$ for EID-CT.

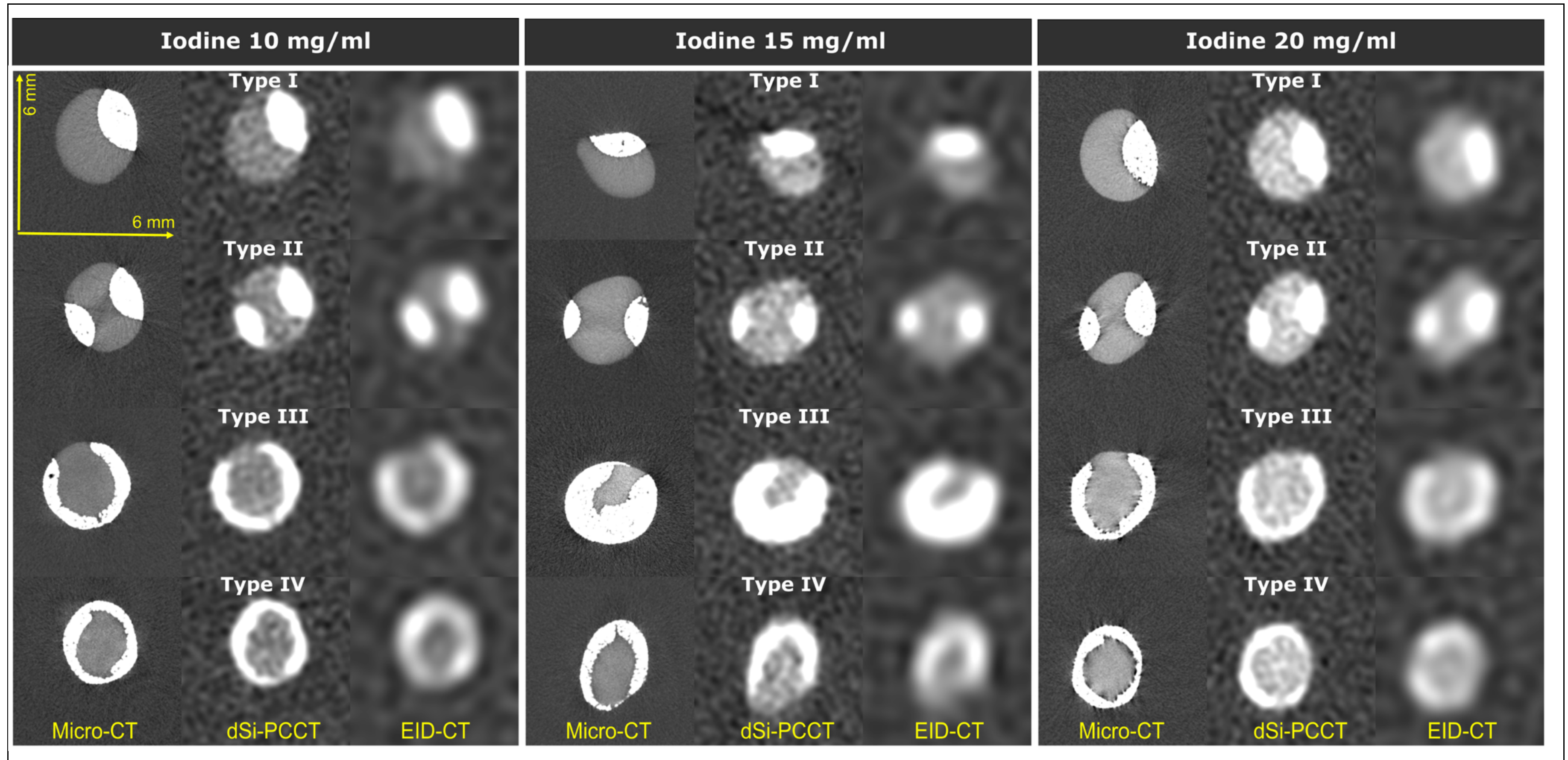

**Figure 4:** Representative images from three CT systems at the slice of maximum calcification for Type I–IV calcification geometries across three iodine concentrations. Images are displayed with a window width of 1000 HU and window level of 300 HU.

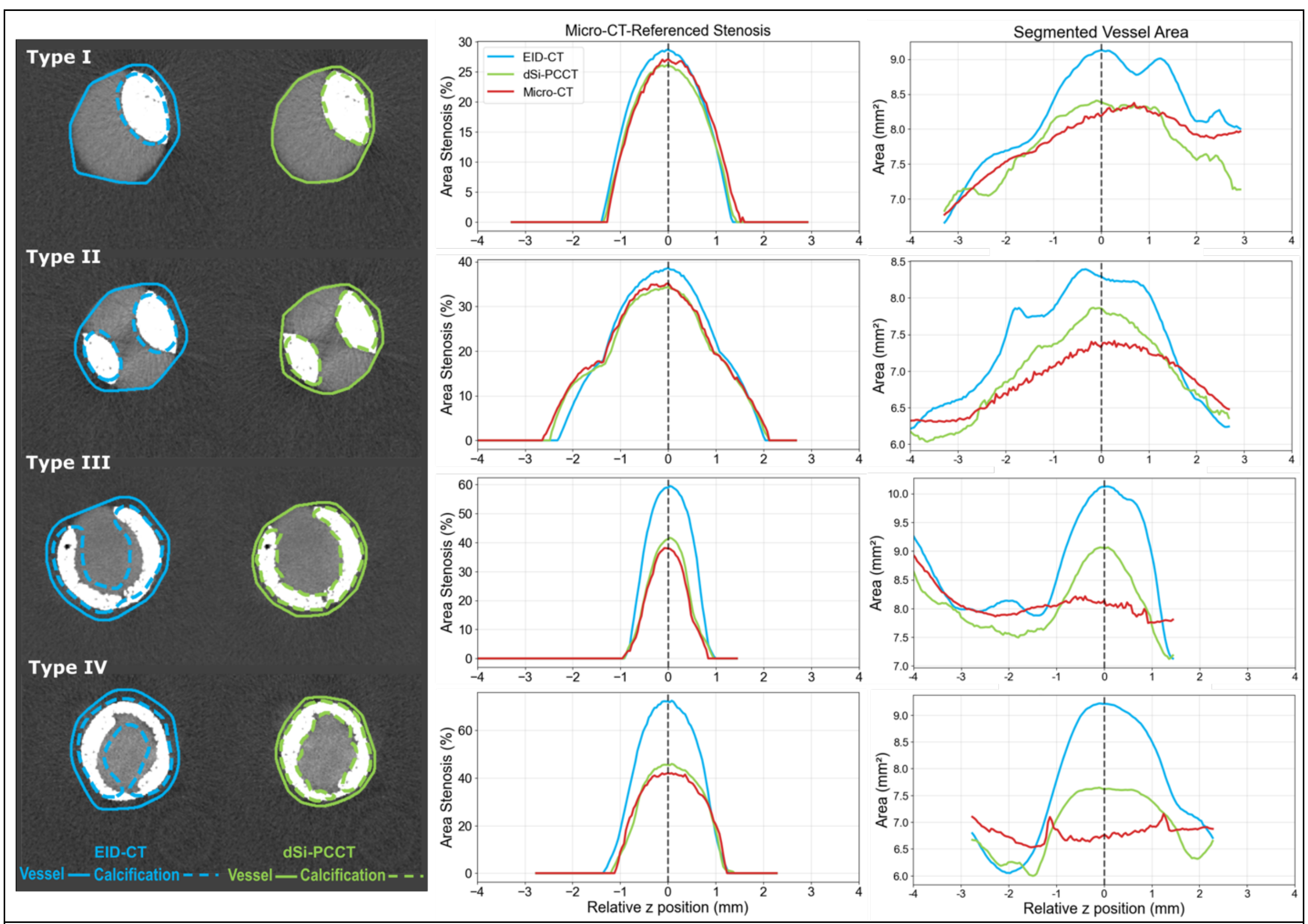

**Figure 5:** (Iodine 10 mg/ml) Representative Micro-CT images of Type I–IV calcification geometries at the slice of maximum calcification, overlaid with outlines of the segmented vessel contour (solid line) and calcification mask (dashed line) for EID-CT and dSi-PCCT. The last two columns show Micro-CT-referenced percent area stenosis and segmented vessel-area profiles along the longitudinal (z) direction for each calcification type. Profiles were aligned at the modality-specific slice of maximum calcification, defined as $z = 0$, to demonstrate longitudinal variation in vessel geometry and blooming-related measurement error across each calcified segment.

Figures 6 and 7 show representative segmentations and corresponding longitudinal profiles for iodine concentrations of 15 and 20 mg/mL, respectively. At 15 mg/mL, dSi-PCCT demonstrated lower profile-level MAE in Micro-CT-referenced percent area stenosis than EID-CT, with values of 0.9%, 2.0%, 4.0%, and 2.0% for Type I–IV calcifications, respectively. The corresponding EID-CT values were 1.2%, 2.7%, 4.3%, and 5.4%.

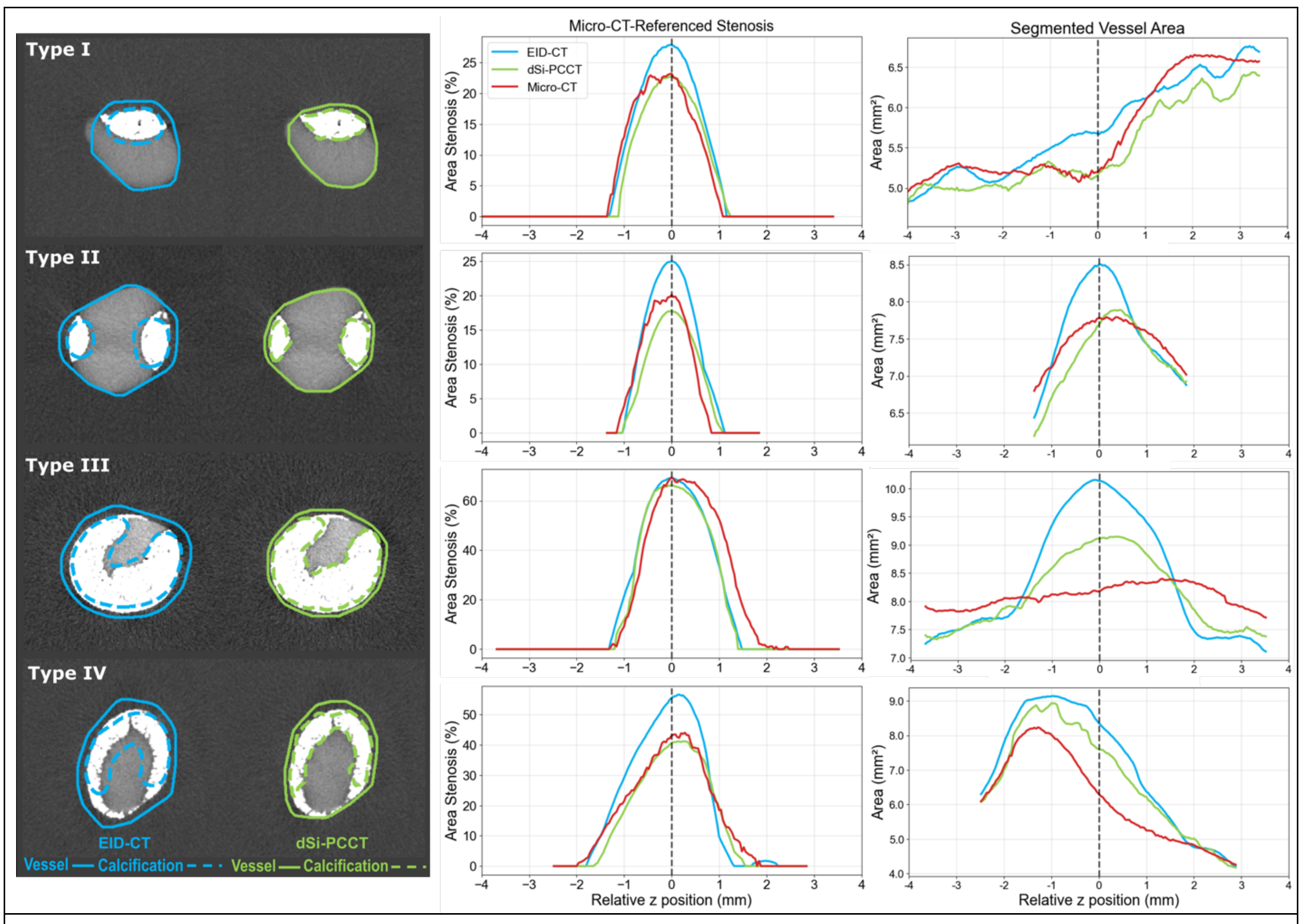

**Figure 6:** (Iodine 15 mg/ml) Representative Micro-CT images of Type I–IV calcification geometries at the slice of maximum calcification, overlaid with outlines of the segmented vessel contour (solid line) and calcification mask (dashed line) for EID-CT and dSi-PCCT. The last two columns show Micro-CT-referenced percent area stenosis and segmented vessel-area profiles along the longitudinal (z) direction for each calcification type. Profiles were aligned at the modality-specific slice of maximum calcification, defined as $z = 0$, to demonstrate longitudinal variation in vessel geometry and blooming-related measurement error across each calcified segment.

For segmented vessel area, dSi-PCCT also demonstrated lower MAE for Type II, Type III, and Type IV calcifications at 15 mg/mL, with errors ranging from 0.24 to 0.63 mm² compared with 0.33 to 0.96 mm² for EID-CT. Differences were smallest for Type I calcifications, with a 0.03 mm² difference between modalities. At 20 mg/mL, profile-level MAE in Micro-CT-referenced percent area stenosis was more comparable between modalities, ranging from 1.42% to 3.47% for dSi-PCCT and from 0.55% to 3.20% for EID-CT. dSi-PCCT nevertheless maintained lower segmented vessel-area MAE, ranging from 0.14 to 0.31 mm² compared with 0.28 to 0.57 mm² for EID-CT.

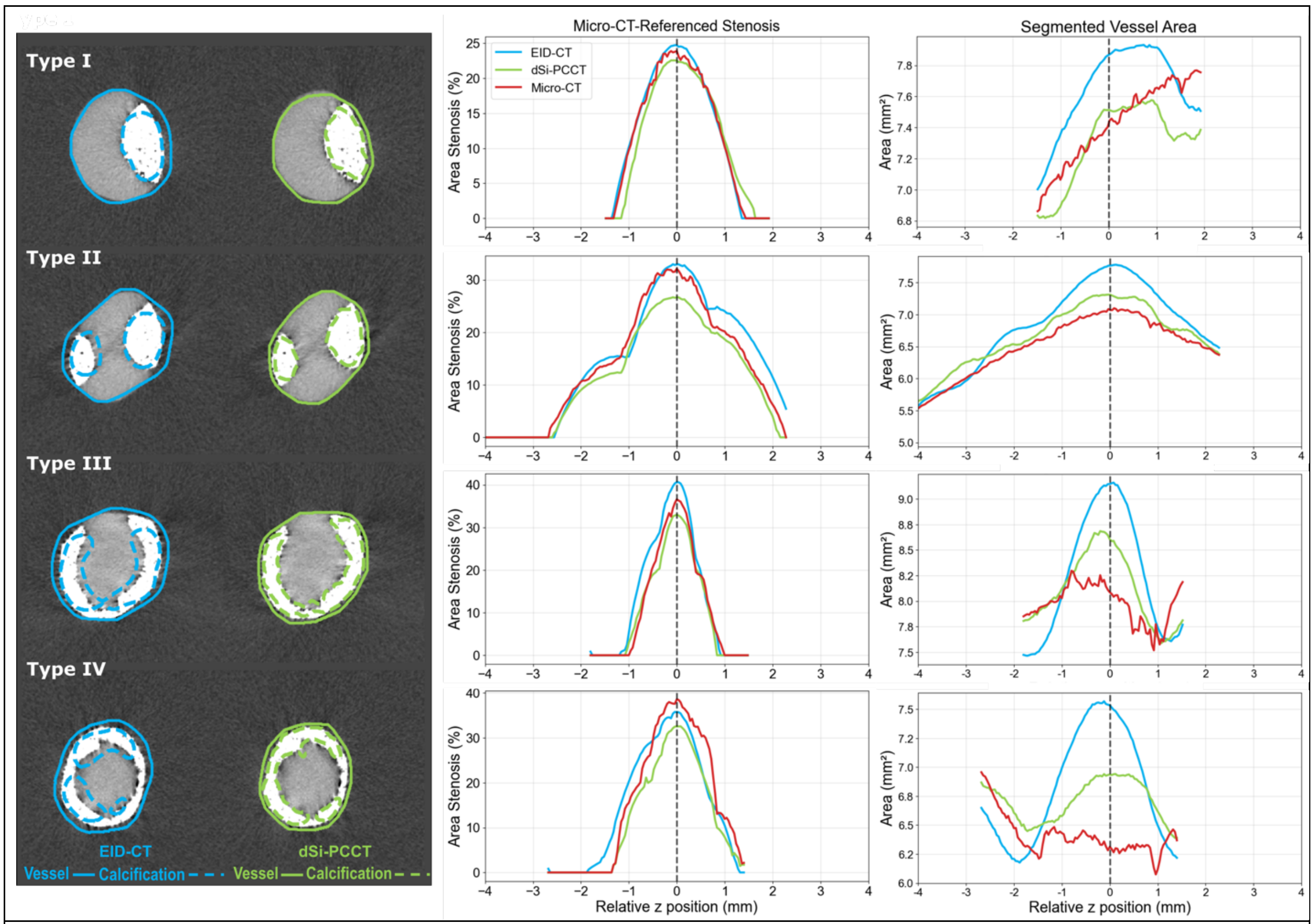

**Figure 7:** (Iodine 20 mg/ml) Representative Micro-CT images of Type I–IV calcification geometries at the slice of maximum calcification, overlaid with outlines of the segmented vessel contour (solid line) and calcification mask (dashed line) for EID-CT and dSi-PCCT. The last two columns show Micro-CT-referenced percent area stenosis and segmented vessel-area profiles along the longitudinal (z) direction for each calcification type. Profiles were aligned at the modality-specific slice of maximum calcification, defined as $z = 0$, to demonstrate longitudinal variation in vessel geometry and blooming-related measurement error across each calcified segment.

Figure 8 summarizes profile-level MAE in Micro-CT-referenced percent area stenosis and segmented vessel area across all calcification geometries and iodine concentrations. Across all evaluated conditions, dSi-PCCT demonstrated lower overall MAE in Micro-CT-referenced percent area stenosis than EID-CT (1.62% versus 3.10%). Mean absolute error in segmented vessel area was also lower overall for dSi-PCCT than for EID-CT (0.31 versus 0.55 mm²). Differences between modalities were statistically significant for Micro-CT-referenced percent area stenosis MAE ($p = 0.027$) and segmented vessel-area MAE ($p = 0.001$).

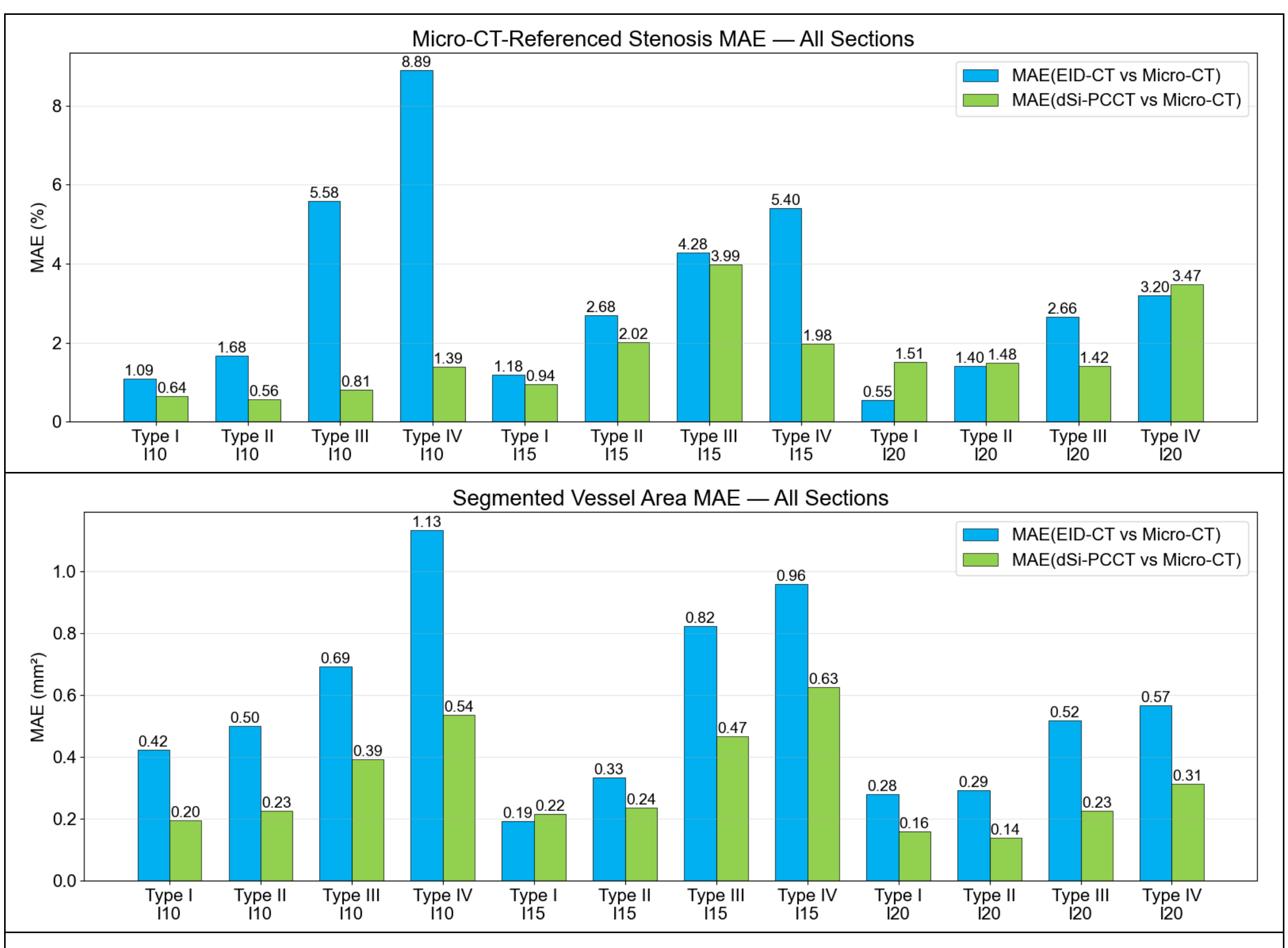

**Figure 8:** Mean absolute error (MAE) of Micro-CT-referenced percent area stenosis and segmented vessel-area profiles for EID-CT and dSi-PCCT across calcification geometries and iodine concentrations (I10 = 10 mg/mL, I15 = 15 mg/mL, and I20 = 20 mg/mL).

*3.2. Secondary Ellipse-Derived Analysis*

Ellipse-derived percent area stenosis and percent vessel-area deviation at the maximum-calcification cross-section are summarized in Table 2. Absolute deviations in ellipse-derived percent area stenosis from the corresponding Micro-CT measurements ranged from 0.1% to 6.5% for dSi-PCCT and from 0.8% to 24.2% for EID-CT across all calcification geometries and iodine concentrations. When averaged across calcification geometries, EID-CT demonstrated deviations ranging from 6.3% at 20 mg/mL to 13.2% at 10 mg/mL, whereas dSi-PCCT deviations ranged from 1.1% at 20 mg/mL to 2.7% at 15 mg/mL. The greatest performance difference was observed at 10 mg/mL, where average deviation decreased from 13.2% with EID-CT to 2.4% with dSi-PCCT. At 20 mg/mL, average deviation decreased from 6.3% with EID-CT to 1.1% with dSi-PCCT.

When analyzed by calcification geometry, absolute deviation in ellipse-derived percent area stenosis increased progressively with calcification complexity for EID-CT. Average deviation across iodine concentrations increased from 3.7% for Type I to 16.9% for Type IV calcifications, with intermediate values of 8.0% and 12.3% for Type II and Type III geometries, respectively. In contrast, dSi-PCCT demonstrated average deviations of 1.8%, 1.4%, 1.1%, and 3.9% for Type I–IV geometries, respectively. The largest difference between modalities occurred for Type IV calcifications, for which dSi-PCCT reduced average deviation by 13.0 percentage points. Type I calcifications demonstrated the smallest difference, at 1.8 percentage points.

Using Micro-CT as the reference, mean absolute differences in percent vessel-area deviation were lower for dSi-PCCT than for EID-CT across all iodine concentrations and calcification geometries (15.0% versus 26.6%). When analyzed by iodine concentration, mean absolute differences for EID-CT were 20.1%, 32.0%, and 27.8% at 10, 15, and 20 mg/mL, respectively, compared with 9.6%, 20.7%, and 14.8% for dSi-PCCT. When analyzed by calcification geometry, mean absolute differences for EID-CT were 13.6%, 27.1%, 31.1%, and 34.7% for Type I–IV calcifications, respectively. The corresponding dSi-PCCT values were 7.4%, 16.6%, 15.1%, and 21.1%.

The largest difference between modalities was observed for Type III calcifications and at 20 mg/mL iodine, whereas the smallest differences were observed for Type I calcifications and at 10 mg/mL iodine. Absolute differences from the corresponding Micro-CT measurements were significantly lower with dSi-PCCT than with EID-CT for ellipse-derived percent area stenosis and percent vessel-area deviation ( $p < 0.001$).

**Table 2:** Ellipse-derived percent area stenosis and percent vessel-area deviation at the maximum-calcification cross-section for Micro-CT, dSi-PCCT, and EID-CT across calcification geometries and iodine concentrations. Bold indicates the more accurate measurement between EID-CT and dSi-PCCT.

| Ellipse-Derived Percent Area Stenosis | | | | | | | | | |
|---|---|---|---|---|---|---|---|---|---|
| | Iodine 10 mg/ml | | | Iodine 15 mg/ml | | | Iodine 20 mg/ml | | |
| | Micro-CT | dSi-PCCT | EID-CT | Micro-CT | dSi-PCCT | EID-CT | Micro-CT | dSi-PCCT | EID-CT |
| Type I | 26.96 | **26.02** | 28.20 | 23.13 | **26.79** | 32.05 | 24.01 | 23.12 | **24.84** |
| Type II | 35.90 | **38.37** | 45.86 | 20.75 | **20.84** | 27.71 | 33.44 | **31.77** | 40.60 |
| Type III | 38.86 | **41.45** | 56.36 | 74.31 | **73.59** | 79.21 | 37.79 | **37.91** | 52.26 |
| Type IV | 41.37 | **44.90** | 65.56 | 41.01 | **47.46** | 64.80 | 37.48 | **35.63** | 40.28 |
| % Deviation in Vessel Area | | | | | | | | | |
| | Iodine 10 mg/ml | | | Iodine 15 mg/ml | | | Iodine 20 mg/ml | | |
| | Micro-CT | dSi-PCCT | EID-CT | Micro-CT | dSi-PCCT | EID-CT | Micro-CT | dSi-PCCT | EID-CT |
| Type I | 0.52 | **2.15** | 9.70 | 0.31 | **17.13** | 25.62 | 0.48 | **4.12** | 6.66 |
| Type II | 2.13 | **19.85** | 34.52 | 3.36 | **16.38** | 21.19 | 4.26 | **23.31** | 35.34 |
| Type III | 2.51 | **11.98** | 18.89 | 7.08 | **23.72** | 41.78 | 3.32 | **22.41** | 45.55 |
| Type IV | 1.53 | **11.28** | 23.82 | 4.77 | **41.06** | 54.89 | 2.89 | **20.24** | 34.61 |

## 4. DISCUSSION

In this study, we evaluated the impact of high-resolution dSi-PCCT on coronary stenosis quantification using anatomically realistic calcified coronary artery phantoms. In the primary Micro-CT-referenced analysis, dSi-PCCT reduced whole-profile mean absolute error in percent area stenosis from 3.10% with EID-CT to 1.62% (p=0.027). dSi-PCCT also reduced mean absolute error in segmented vessel area from 0.55 to 0.31 mm² (p=0.001). These improvements were observed across the evaluated calcification morphologies and iodine concentrations and were most pronounced for extensively calcified vessels. Together, these findings indicate that the higher-resolution dSi-PCCT protocol provided more accurate recovery of calcification burden and vessel geometry along the full calcified vessel segment.

The secondary ellipse-derived analysis supported the same overall conclusion. At the maximum-calcification cross-section, dSi-PCCT produced ellipse-derived percent area stenosis values that were consistently closer to the corresponding Micro-CT measurements than those obtained with EID-CT. Percent vessel-area deviation relative to the ellipse-estimated reference was also lower with dSi-PCCT than with EID-CT (15.0% versus 26.6%). The agreement between the primary and secondary findings suggests that the observed advantage of dSi-PCCT was not dependent on a single reference strategy. However, the ellipse-derived reference should be interpreted as a local geometric surrogate intended for settings without an independent ground truth, rather than as a replacement for the Micro-CT reference.

Prior phantom studies with known ground truths have demonstrated that improved spatial resolution and reduced blooming can enhance the visualization and quantification of coronary stents, stenoses, and calcified plaques (Koons *et al.*, 2022; Verelst *et al.*, 2023). Our findings with dSi-PCCT are consistent with these observations. The lower errors in segmented vessel area indicate that dSi-PCCT reduced the apparent expansion of the vessel boundary associated with calcium blooming. This improvement was accompanied by lower errors in percent area stenosis, supporting the relationship between more accurate vessel and calcification delineation and improved stenosis quantification.

Calcification morphology influenced the magnitude of the observed improvement. Previous Micro-CT and ex vivo coronary studies have shown that blooming-related measurement errors depend on calcium shape, extent, and spatial distribution (Sandstedt *et al.*, 2021; Marsh *et al.*, 2023). Sandstedt et al. observed that ring-shaped calcifications exhibited larger measurement discrepancies than more localized calcifications, suggesting that calcium morphology and quantity influence the magnitude and spatial distribution of blooming artifacts (Sandstedt *et al.*, 2021). In the present study, Type III and Type IV calcifications, which occupied three to four quadrants of the vessel circumference, produced larger vessel-area and percent area stenosis errors with EID-CT than the more localized Type I and Type II calcifications. These observations are consistent with prior findings that extensive and ring-like calcifications are particularly susceptible to partial-volume-related blurring (Sandstedt *et al.*, 2021). Importantly, dSi-PCCT reduced these morphology-dependent errors and provided more accurate measurements across the range of calcification geometries.

We also examined the effect of luminal contrast level on the advantages of high-resolution imaging. For the acquisition and reconstruction parameters used in this study, the three iodine concentrations (10, 15, and 20 mg/ml) produced mean luminal attenuations of approximately 276, 364, and 435 HU for EID-CT and 337, 455, and 553 HU for dSi-PCCT. All tested conditions fell within the range relevant to coronary stenosis assessment: each exceeded the 200 HU threshold reported as adequate for accurate low-density stenosis quantification in vessels of at least 3 mm (Toepker *et al.*, 2013), and together they spanned the 350 HU level reported as optimal for coronary stenosis detection (Fei *et al.*, 2008). The largest improvement with dSi-PCCT was observed at the lowest iodine concentration. Under lower-enhancement conditions, reduced separation between lumen and calcification attenuation increases the effect of partial-volume blurring on the measured boundaries, thereby increasing the relative benefit of improved spatial resolution. Nevertheless, the dSi-PCCT advantage was not limited to the lowest iodine concentration and remained evident across the evaluated attenuation range.

The Micro-CT-referenced percent area stenosis metric used in this study differs from conventional quantitative CCTA area stenosis, which is calculated from the minimum lumen area relative to an averaged or interpolated reference lumen area from nearby normal vessel segments (Nieman *et al.*, 2024). Because the patient-derived phantom vessels varied in cross-sectional size along the longitudinal direction, applying a fixed proximal or distal reference could confound true vessel-caliber variation with blooming-related error. The primary analysis therefore used the corresponding Micro-CT vessel area at each longitudinal position and evaluated the complete percent area stenosis and segmented vessel-area profiles, whereas the secondary analysis used a local ellipse-estimated reference at the maximum-calcification cross-section. Accordingly, these measurements should be interpreted as task-based metrics of imaging accuracy rather than as conventional clinical area stenosis or mapped directly to clinical stenosis categories.

Several additional limitations should be considered. First, the evaluation was performed using coronary artery phantoms under static imaging conditions. This design enabled controlled comparison of the imaging systems against known ground truth while excluding cardiac motion as a confounding factor. However, clinical translation will require adequate cardiac gating and temporal resolution. Future dynamic phantom and in vivo studies should therefore evaluate the combined effects of cardiac motion and calcium blooming, as motion-induced blurring may reduce the benefits associated with improved intrinsic spatial resolution.

Second, reconstruction protocols were system-specific but selected to maximize spatial resolution for this calcification task. EID-CT images were reconstructed using ASiR-V 50% with a Bone kernel, whereas dSi-PCCT images were reconstructed using DL-High in UHD Ultra mode with a smaller voxel size, a $1024 \times 1024$ matrix, and a 12-cm field of view. These resolution-optimized settings are consistent with Society of Cardiovascular CT (SCCT) recommendations to use thin slices, a small field of view, a large matrix, and sharper kernels for quantitative coronary assessment, although reconstruction settings also affect image noise and quantitative measurements (Achenbach *et al.*, 2010; Nieman *et al.*, 2024). Thus, this study compared the high-resolution task performance achievable with each system under matched acquisition and dose conditions. Future studies should also compare routine and high-resolution clinical reconstructions.

Third, the analysis relied on a fully automated threshold-based segmentation pipeline. This approach provided consistent processing across modalities and reduced operator dependence, but it does not reproduce the clinical workflow in which vessel boundaries and stenosis severity are assessed by readers. Although the visibly tighter dSi-PCCT contours are consistent with reduced blooming, reader-based studies and independent segmentation validation would provide complementary evidence that the observed differences reflect improved boundary fidelity rather than modality-dependent behavior of the segmentation method.

Finally, this study did not use the spectral capabilities of dSi-PCCT, including material decomposition and virtual monoenergetic imaging. Spectral separation of iodine and calcium could provide additional improvements in lumen and calcification delineation and warrants evaluation in future work. The study also evaluated a single dSi-PCCT system, dose level, and reconstruction protocol. Further studies are needed to determine the extent to which these findings generalize across detector technologies, reconstruction settings, patient sizes, dynamic imaging conditions, and clinical CCTA examinations.

## 5. CONCLUSIONS

In summary, silicon-based photon-counting CT reduced blooming artifacts and improved coronary stenosis quantification compared with EID-CT across a range of calcification morphologies and luminal attenuation levels. The greatest improvements were observed for heavily calcified vessels, where blooming-related errors are most pronounced. These findings suggest that the improved spatial resolution of dSi-PCCT has the potential to enhance the noninvasive assessment of calcified coronary stenosis.

## ACKNOWLEDGMENTS

This work was supported by GE HealthCare and Stanford Propel Postdoctoral Fellowship. The development of the vessel phantom reported in this publication was supported by the National Heart, Lung, and Blood Institute of the National Institutes of Health under Award Number R01 HL151561. The content of this publication is solely the responsibility of the authors and does not necessarily represent the official views of the National Institutes of Health.

## REFERENCES

Achenbach, S. *et al.* (2010) “Influence of slice thickness and reconstruction kernel on the computed tomographic attenuation of coronary atherosclerotic plaque,” *Journal of Cardiovascular Computed Tomography*, 4(2), pp. 110–115. Available at: https://doi.org/10.1016/j.jcct.2010.01.013.

Boccalini, S. *et al.* (2022) “First In-Human Results of Computed Tomography Angiography for Coronary Stent Assessment With a Spectral Photon Counting Computed Tomography.,” *Investigative radiology*, 57(4), pp. 212–221. Available at: https://doi.org/10.1097/RLI.0000000000000835.

Boussoussou, M. *et al.* (2025) “Comparative analysis of photon-counting and energy-integrating detector CT to identify obstructive coronary artery disease,” *European Radiology*, 36(4), pp. 3091–3102. Available at: https://doi.org/10.1007/s00330-025-12118-7.

Fahrni, G. *et al.* (2025) “Quantification of Coronary Artery Stenosis in Very-High-Risk Patients Using Ultra-High Resolution Spectral Photon-Counting CT,” *Investigative Radiology*, 60(2), pp. 114–122. Available at: https://doi.org/10.1097/RLI.0000000000001109.

Fei, X. *et al.* (2008) “64-MDCT Coronary Angiography: Phantom Study of Effects of Vascular Attenuation on Detection of Coronary Stenosis,” *American Journal of Roentgenology*, 191(1), pp. 43–49. Available at: https://doi.org/10.2214/AJR.07.2653.

Gulati, M. *et al.* (2021) "2021 AHA/ACC/ASE/CHEST/SAEM/SCCT/SCMR Guideline for the Evaluation and Diagnosis of Chest Pain: A Report of the American College of Cardiology/American Heart Association Joint Committee on Clinical Practice Guidelines," *Circulation*, 144(22). Available at: https://doi.org/10.1161/CIR.0000000000001029.

Hagar, M.T. *et al.* (2023) "Accuracy of Ultrahigh-Resolution Photon-counting CT for Detecting Coronary Artery Disease in a High-Risk Population," *Radiology*, 307(5). Available at: https://doi.org/10.1148/radiol.223305.

Holmes, T.W. *et al.* (2024) "Ultrahigh-Resolution K-Edge Imaging of Coronary Arteries With Prototype Deep-Silicon Photon-Counting CT: Initial Results in Phantoms," *Radiology*, 311(3). Available at: https://doi.org/10.1148/radiol.231598.

Kalisz, K. *et al.* (2016) "Artifacts at Cardiac CT: Physics and Solutions," *RadioGraphics*, 36(7), pp. 2064–2083. Available at: https://doi.org/10.1148/rg.2016160079.

Koons, E. *et al.* (2022) "Improved quantification of coronary artery luminal stenosis in the presence of heavy calcifications using photon-counting detector CT.," *Proceedings of SPIE--the International Society for Optical Engineering*, 12031. Available at: https://doi.org/10.1117/12.2613019.

Koons, E.K. *et al.* (2024) "Coronary artery stenosis quantification in patients with dense calcifications using ultra-high-resolution photon-counting-detector computed tomography," *Journal of Cardiovascular Computed Tomography*, 18(1), pp. 56–61. Available at: https://doi.org/10.1016/j.jcct.2023.10.009.

Kotronias, R.A. *et al.* (2025) "Benchmarking Photon-Counting Computed Tomography Angiography Against Invasive Assessment of Coronary Stenosis," *JACC: Cardiovascular Imaging*, 18(5), pp. 572–585. Available at: https://doi.org/10.1016/j.jcmg.2024.11.005.

Latina, J. *et al.* (2021) "Ultra-High-Resolution Coronary CT Angiography for Assessment of Patients with Severe Coronary Artery Calcification: Initial Experience," *Radiology: Cardiothoracic Imaging*, 3(4), p. e210053. Available at: https://doi.org/10.1148/ryct.2021210053.

Lowekamp, B.C. *et al.* (2013) "The Design of SimpleITK," *Frontiers in Neuroinformatics*, 7. Available at: https://doi.org/10.3389/fninf.2013.00045.

Marsh, J.F. *et al.* (2023) "Ex vivo coronary calcium volume quantification using a high-spatial-resolution clinical photon-counting-detector computed tomography," *Journal of Medical Imaging*, 10(04). Available at: https://doi.org/10.1117/1.JMI.10.4.043501.

Mergen, V. *et al.* (2022) "Ultra-High-Resolution Coronary CT Angiography With Photon-Counting Detector CT," *Investigative Radiology*, 57(12), pp. 780–788. Available at: https://doi.org/10.1097/RLI.0000000000000897.

Narula, J. *et al.* (2021) "SCCT 2021 Expert Consensus Document on Coronary Computed Tomographic Angiography: A Report of the Society of Cardiovascular Computed Tomography.,"

*Journal of cardiovascular computed tomography*, 15(3), pp. 192–217. Available at: https://doi.org/10.1016/j.jcct.2020.11.001.

Nieman, K. *et al.* (2024) "Standards for quantitative assessments by coronary computed tomography angiography (CCTA)," *Journal of Cardiovascular Computed Tomography*, 18(5), pp. 429–443. Available at: https://doi.org/10.1016/j.jcct.2024.05.232.

Pack, J.D. *et al.* (2024) "A coronary artery phantom for task-based CT performance assessment and a comparative study of clinical CT, photon counting CT, and micro CT," *arXiv preprint arXiv:2401.04215* [Preprint].

Qi, L. *et al.* (2016) "The Diagnostic Performance of Coronary CT Angiography for the Assessment of Coronary Stenosis in Calcified Plaque.," *PloS one*, 11(5), p. e0154852. Available at: https://doi.org/10.1371/journal.pone.0154852.

Sandstedt, M. *et al.* (2021) "Improved coronary calcification quantification using photon-counting-detector CT: an ex vivo study in cadaveric specimens," *European Radiology*, 31(9), pp. 6621–6630. Available at: https://doi.org/10.1007/s00330-021-07780-6.

Song, Y. Bin *et al.* (2019) "Contemporary Discrepancies of Stenosis Assessment by Computed Tomography and Invasive Coronary Angiography," *Circulation: Cardiovascular Imaging*, 12(2). Available at: https://doi.org/10.1161/CIRCIMAGING.118.007720.

Toepker, M. *et al.* (2013) "Stenosis Quantification of Coronary Arteries in Coronary Vessel Phantoms With Second-Generation Dual-Source CT: Influence of Measurement Parameters and Limitations," *American Journal of Roentgenology*, 201(2), pp. W227–W234. Available at: https://doi.org/10.2214/AJR.12.9453.

Verelst, E. *et al.* (2023) "Stent appearance in a novel silicon-based photon-counting CT prototype: ex vivo phantom study in head-to-head comparison with conventional energy-integrating CT," *European Radiology Experimental*, 7(1). Available at: https://doi.org/10.1186/s41747-023-00333-0.

Verelst, E. *et al.* (2025) "Plaque and stent distinction using an experimental silicon-based photon-counting CT: An experimental and clinical comparison with conventional energy-integrated detector CT," *European Journal of Radiology*, 189. Available at: https://doi.org/10.1016/j.ejrad.2025.112207.